\documentclass{article}

\usepackage[utf8]{inputenc}
\usepackage[T1]{fontenc}
\usepackage{amsmath,amssymb,amsfonts,amsthm}
\usepackage{graphicx}
\usepackage{booktabs}
\usepackage[hidelinks]{hyperref}
\usepackage{float}
\usepackage{xcolor}
\usepackage[margin=1in]{geometry}
\usepackage{natbib}
\usepackage{subcaption}
\usepackage{tikz}
\usepackage{pgfplots}
\pgfplotsset{compat=1.17}
\usetikzlibrary{positioning,arrows.meta,calc,fit,backgrounds,shapes.geometric}

\newcommand{\E}{\mathbb{E}}
\newcommand{\model}{\textsc{BF-ConvUNeXt}}
\newcommand{\convnext}{\textsc{ConvNeXt}}
\newcommand{\bfbn}{\textsc{BiasFreeBatchNorm}}
\newcommand{\sigpx}{\sigma_{255}}

\definecolor{oursblue}{RGB}{31,119,180}
\definecolor{classicgray}{RGB}{127,127,127}
\definecolor{sotaorange}{RGB}{255,127,14}
\definecolor{gaborgreen}{RGB}{44,160,44}
\definecolor{skipred}{RGB}{214,39,40}
\definecolor{bottlepurple}{RGB}{148,103,189}

\title{Small, Bias-Free, Blind and Convolutional Denoiser\\
\large{A compact ConvNeXt U-Net for blind Gaussian color-image denoising}}

\author{
  Nikolas Markou \\
  \texttt{nikolas.markou@electiconsulting.com}
}

\date{}

\begin{document}

\maketitle

% ============================================================================
\begin{abstract}
We describe and evaluate \model{}, a compact bias-free \convnext{} U-Net for blind additive-white-Gaussian-noise (AWGN) color image denoising. It combines four ingredients, none of them new, so that a single property survives end to end. A frozen depthwise Gabor stem gives an oriented band-pass front end with zero trainable parameters. A Laplacian-pyramid encoder routes the full-resolution high-frequency residual into each skip connection, with three extra \convnext{} blocks per band. A \convnext{}-V1 U-Net forms the body. The construction is bias-free throughout: no additive bias in any convolution, a linear head, LeakyReLU activations, and a variance-only batch norm. Together these make the $0.82$M-parameter network exactly degree-1 homogeneous at inference, $D(\alpha y)=\alpha D(y)$, which licenses the Miyasawa/Tweedie reading of the residual as a score and, following Mohan et al., supports blind generalization across noise levels from one model. We train a single blind model on a noise-$\sigma$ curriculum spanning $\sigpx\!\approx\!6.4$ to $64$; it extrapolates far past that ceiling without a cliff, degrading smoothly to $22.8$\,dB at $\sigpx{=}150$ and $20.0$\,dB at $\sigpx{=}200$ ($2.3\times$ and $3.1\times$ the training maximum). Evaluated unchanged on the four standard color sets (CBSD68, Kodak24, McMaster, Urban100) at $\sigma\!\in\!\{15,25,50\}$ under the full-image protocol, it matches or beats DnCNN and FFDNet on every set and level, averaging about $+0.7$\,dB over DnCNN while FFDNet, though non-blind, still trails. The closest case is CBSD68: there the $+0.10$ to $+0.15$\,dB margin over DnCNN falls within the sensitivity of our disclosed $[0,1]$ input clip, so the result is best read as a match (Section~\ref{sec:limitations}). Against the heavyweight CNN/transformer state of the art it sits a small, consistent margin behind ($\sim0.3$--$0.5$\,dB on CBSD68, $\sim0.3$--$0.7$\,dB on Kodak24, $\sim0.5$--$0.9$\,dB on McMaster, and $\sim0.6$--$1.7$\,dB on Urban100), at roughly $1/15$ to $1/39$ of their parameters. The homogeneity is inference-only and checkpoint-specific, and the learned residual is a \emph{local} score, not a global prior (its Jacobian is non-conservative), so plug-and-play/RED and calibrated-posterior guarantees do not transfer. That same local score still drives stochastic sampling from the implicit prior and a range of linear inverse problems (inpainting, super-resolution, deblurring, compressive sensing). Results come from a single checkpoint, reported in PSNR (plus SSIM for our own model), without ablations.
\end{abstract}

% ============================================================================
\section{Introduction}
\label{sec:intro}

Blind Gaussian image denoising recovers a clean image $x$ from $y = x + \varepsilon$, $\varepsilon \sim \mathcal{N}(0,\sigma^2 I)$, with $\sigma$ unknown at test time. It is both a canonical low-level vision task and a common building block inside modern inverse-problem solvers. This paper is framed by two facts. The field's accuracy is dominated by large transformer/CNN hybrids (DRUNet, SwinIR, Restormer, SCUNet) with tens of millions of parameters, often trained per-noise-level. Set against them is a much older and simpler observation, still underexploited in a modern architecture: removing all additive constants from a denoising CNN makes it exactly scale-equivariant, and therefore able to generalize blindly across noise levels that a biased network cannot handle \citep{mohan2020biasfree}.

This paper takes the second fact seriously and asks one empirical question: how far does a compact, strictly bias-free, modern-conv-block U-Net get on the standard color-denoising leaderboard as a single blind model? Our answer is \model{}, a $0.82$M-parameter bias-free \convnext{} U-Net that composes four ideas we did not invent:

\begin{itemize}
  \item a frozen depthwise Gabor stem \citep{gabor1946}: a bank of oriented, frequency-tuned band-pass filters fixed at initialization (zero trainable parameters), whose outputs are remixed into the stem channels by a single trainable bias-free $1{\times}1$ projection;
  \item a Laplacian-pyramid encoder \citep{burt1983laplacian} that, at each downsampling junction, sends the full-resolution high-frequency residual band to the skip connection and passes the low-frequency band downward, with three extra \convnext{} blocks applied to each high-frequency skip band;
  \item a \convnext{}-V1 U-Net body \citep{liu2022convnext,ronneberger2015unet} (depthwise $7{\times}7$ conv, inverted-bottleneck MLP, LayerScale);
  \item bias-freedom from end to end: no additive bias in any layer, a linear head, LeakyReLU activations, and a variance-only batch normalization, chosen together so the network is globally degree-1 homogeneous at inference.
\end{itemize}

The contribution is the specific, bias-preserving way they are combined so that homogeneity survives end to end, together with an apples-to-apples measurement against the published leaderboard on the same test sets and full-image protocol. This is an integration-and-evaluation paper, closer to an engineering report than to a novel-component paper; it makes three contributions:

\begin{enumerate}
  \item a precise description of \model{} (topology, frozen Gabor stem, Laplacian-pyramid skip decomposition, \convnext{}-V1 block, and the bias-free construction that makes the network degree-1 homogeneous), at a level of detail sufficient to reimplement it;
  \item a fair benchmark: one blind checkpoint, evaluated unchanged on all four standard color sets at three noise levels under the full-image protocol, tabulated against six published baselines;
  \item an explicit account of the theory's limits: the Miyasawa/Tweedie residual-as-score reading is only a motivation, and we show (Section~\ref{sec:theory-limits}) that the learned operator is non-conservative, so it is not the gradient of any global prior.
\end{enumerate}

The bias-free construction also reaches beyond denoising. Because the residual is a Miyasawa score, the same frozen model serves as the implicit prior in the stochastic inverse-problem solver of \citet{kadkhodaie2021stochastic}, extending with no retraining to linear inverse problems and to sampling from the implicit prior (Section~\ref{sec:limitations}). This is not an ablation study, and the leaderboard comparison is PSNR-only because the collated source is (we add SSIM for \model{} in Table~\ref{tab:ssim}). The result is favorable: the compact blind model matches or beats the classic sub-1M CNNs (a clear win on Kodak24, McMaster, and Urban100; a match on CBSD68, Section~\ref{sec:limitations}) and lands within about a dB of models an order of magnitude larger.

% ============================================================================
\clearpage
\section{Architecture}
\label{sec:arch}

\begin{figure}[H]
\centering
\resizebox{0.8\textwidth}{!}{%
\begin{tikzpicture}[
  font=\small,
  every node/.style={transform shape},
  box/.style={draw, thick, rounded corners=3pt, minimum width=3.0cm, minimum height=1.35cm,
              text width=2.7cm, align=center, inner sep=5pt},
  stem/.style={box, fill=gaborgreen!15},
  enc/.style={box, fill=oursblue!12},
  dec/.style={box, fill=oursblue!12},
  bott/.style={box, fill=bottlepurple!16},
  io/.style={box, minimum height=1.1cm, text width=2.5cm, fill=black!5},
  skip/.style={draw, thick, rounded corners=3pt, minimum width=2.7cm, minimum height=1.0cm,
               text width=2.5cm, align=center, inner sep=4pt, fill=skipred!12, font=\footnotesize},
  add/.style={draw, thick, circle, minimum size=0.9cm, inner sep=0pt, fill=black!8},
  arr/.style={-{Latex[length=3mm]}, very thick},
  sarr/.style={-{Latex[length=3mm]}, very thick, skipred!80, dashed},
  lbl/.style={font=\footnotesize, inner sep=1.5pt},
  ]
  % --- encoder column (left, top to bottom) ---
  \node[io]   (in)  at (0,0)     {input $y$\\[1pt] {\footnotesize $[0,1]$}};
  \node[stem] (gab) at (0,-2.1)  {frozen Gabor stem\\[1pt] {\footnotesize $3{\times}22{=}66$, $11{\times}11$}};
  \node[enc]  (gabor_stem_projection) at (0,-4.2) {stem projection\\[1pt] {\footnotesize trainable, bias-free\\ $1{\times}1$, $66{\to}66$}};
  \node[enc]  (e0)  at (0,-6.3)  {encoder L0\\[1pt] {\footnotesize $66$ ch, $3\times$ block}};
  \node[enc]  (e1)  at (0,-10.1) {encoder L1\\[1pt] {\footnotesize $66$ ch, $3\times$ block}};
  % --- bottleneck (bottom center) ---
  \node[bott] (bn)  at (4,-11.9) {bottleneck\\[1pt] {\footnotesize $66$ ch, $3\times$ block}};
  % --- skip nodes (high-frequency bands, middle column) ---
  \node[skip] (s0)  at (4,-6.3)  {hi-freq skip L0\\[1pt] $+\,3\times$ block};
  \node[skip] (s1)  at (4,-10.1) {hi-freq skip L1\\[1pt] $+\,3\times$ block};
  % --- decoder spine (right column): Add node -> decoder block, per level ---
  \node[add]  (a1)  at (8,-10.1) {$\oplus$};
  \node[dec]  (d1)  at (8,-8.2)  {decoder L1\\[1pt] {\footnotesize $66$ ch, $3\times$ block}};
  \node[add]  (a0)  at (8,-6.3)  {$\oplus$};
  \node[dec]  (d0)  at (8,-3.2)  {decoder L0\\[1pt] {\footnotesize $66$ ch, $3\times$ block}};
  \node[io]   (out) at (8,0)     {output $\hat{x}=D(y)$\\[1pt] {\footnotesize dense $1{\times}1$}};
  % --- encoder flow (low-frequency band descends) ---
  \draw[arr] (in)  -- (gab);
  \draw[arr] (gab) -- (gabor_stem_projection);
  \draw[arr] (gabor_stem_projection) -- (e0);
  \draw[arr] (e0)  -- node[lbl, right] {Lap.\ split $\downarrow$} (e1);
  \draw[arr] (e1)  -- node[lbl, below left=1pt] {Lap.\ split $\downarrow$} (bn);
  % --- decoder spine: each Add merges (upsampled feature + skip), THEN the block runs ---
  \draw[arr] (bn) -- node[lbl, below right=1pt] {upsample $\times2$} (a1);
  \draw[arr] (a1) -- (d1);
  \draw[arr] (d1) -- node[lbl, right] {upsample $\times2$} (a0);
  \draw[arr] (a0) -- (d0);
  \draw[arr] (d0) -- node[lbl, right] {linear head} (out);
  % --- high-frequency skip connections into the Add nodes (separate merge step) ---
  \draw[sarr] (e0) -- (s0);   \draw[sarr] (s0) -- (a0);
  \draw[sarr] (e1) -- (s1);   \draw[sarr] (s1) -- (a1);
\end{tikzpicture}%
}
\caption{\model{} data flow (channel widths shown). The encoder (left) descends, the decoder spine (right) ascends, and the bottleneck sits at the base. A frozen Gabor stem, followed by a trainable bias-free $1{\times}1$ projection (\texttt{gabor\_stem\_projection}) that remixes its $66$ responses, feeds a two-level \convnext{}-V1 U-Net. Downsampling is a Laplacian-pyramid split: the low-frequency band descends (solid arrows) while the full-resolution high-frequency residual becomes the skip (dashed), passing through three extra bias-free \convnext{} blocks. At each decoder level the skip and the upsampled feature are combined at an explicit parameter-free addition ($\oplus$): the upsampled feature is added directly to the skip (both carry $66$ channels, so their widths already agree), and only then is the sum processed by the decoder block. The head is a bias-free dense $1{\times}1$ convolution. Every operation is degree-1 homogeneous at inference, so $D(\alpha y)=\alpha D(y)$.}
\label{fig:arch}
\end{figure}
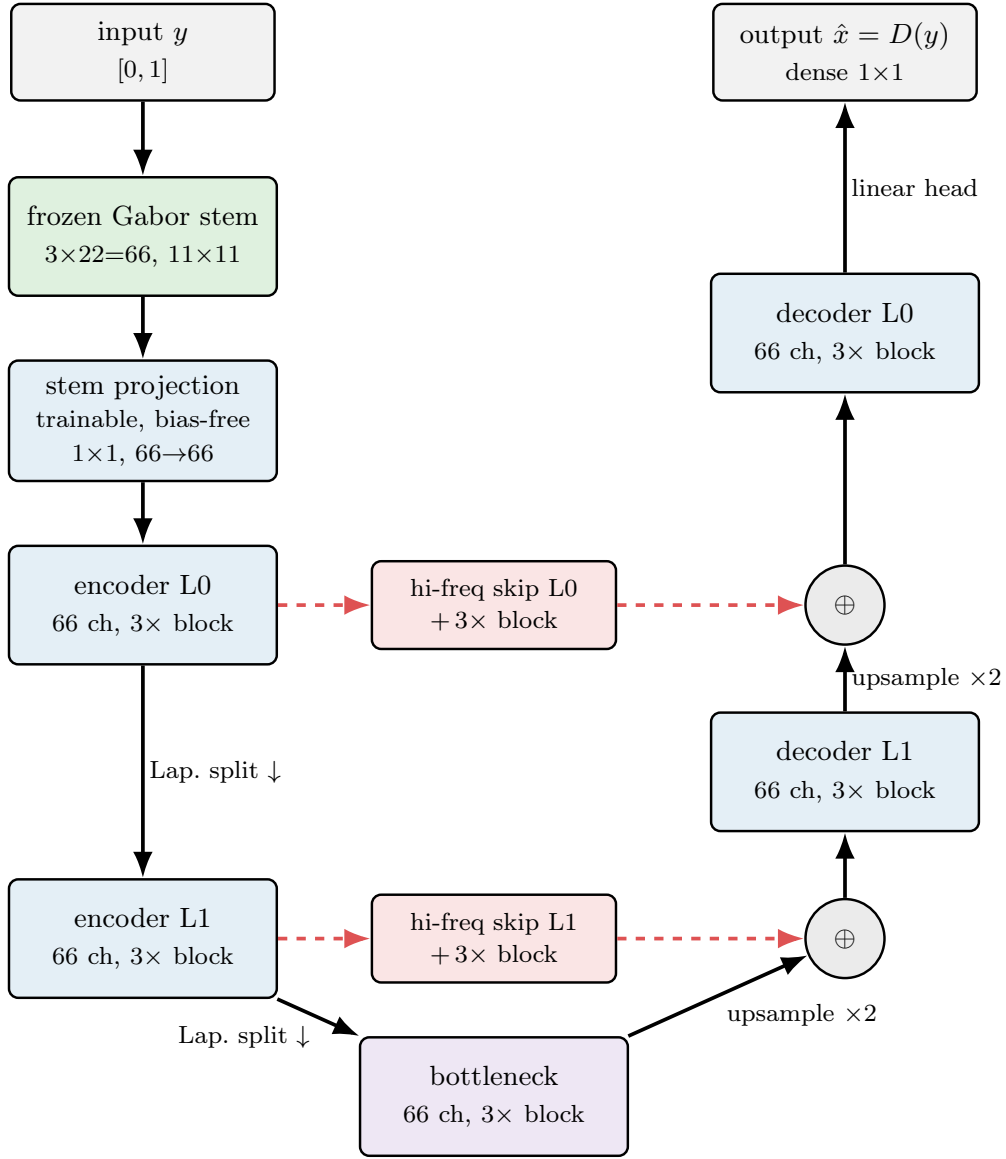

\model{} is a two-level U-Net operating on images normalized to $[0,1]$. All parameter counts and layer shapes below are read from the actual trained checkpoint (\texttt{20260720\_convunext\_denoiser\_hfb3}, $821{,}832$ parameters total: $807{,}048$ trainable and $14{,}784$ frozen). Figure~\ref{fig:arch} shows the data flow.

\subsection{U-Net topology}
Every scope runs at a flat width of $66$ channels: the width multiplier is $1$, so there is no doubling between levels and the two encoder levels, the bottleneck, and the two decoder levels all operate at $66$ channels. Each of the five scopes (two encoder levels, bottleneck, two decoder levels) contains three \convnext{}-V1 blocks. Upsampling is bilinear ($\times 2$). Decoder skips are fused by addition, $\mathrm{dec} \leftarrow \mathrm{Add}(\mathrm{skip},\, \mathrm{up})$, not by concatenation; because every scope runs at the same $66$ channels the skip and upsampled feature already agree in width, so no channel projection is needed. Both choices keep the merge exactly linear and homogeneous.

\subsection{Frozen Gabor stem}
In a plain CNN the first learnable operation on a noisy image is a small learned convolution. Here it is replaced by a fixed oriented band-pass front end. A depthwise convolution with $22$ filters per input channel and an $11{\times}11$ kernel is initialized with a Gabor bank, each output channel being one 2D Gabor wavelet $g(u,v)=\exp\!\big(-\tfrac{u_\theta^2+\gamma^2 v_\theta^2}{2s^2}\big)\cos\!\big(\tfrac{2\pi u_\theta}{\lambda}+\psi\big)$ at a swept orientation, frequency, and phase, and is marked non-trainable, so it contributes no trainable parameters and never adapts. With three input channels it emits $3\times 22 = 66$ feature maps. The rationale is standard signal processing: an oriented multi-scale band-pass decomposition is a strong, noise-agnostic image representation that the network need not spend capacity (or risk overfitting noise) to relearn.

Those $66$ frozen responses are not handed to the encoder as they stand. They first pass through \texttt{gabor\_\allowbreak stem\_\allowbreak projection}, a single trainable bias-free $1{\times}1$ convolution ($66 \to 66$ channels, \texttt{use\_bias}=False, $66 \times 66 = 4{,}356$ parameters), whose output is what we call the $66$ stem channels. The division of labor is deliberate: the bank itself is a fixed, hand-designed dictionary that cannot adapt, while the projection is the one learned degree of freedom on top of it, free to reweight and combine band-pass responses across orientations, frequencies, phases, and the three input color channels without altering a single filter. Being a bias-free linear map it is degree-1 homogeneous and so leaves the property of Section~\ref{sec:biasfree} intact. It is cheap in parameters and less cheap in compute: $4{,}356$ parameters is $0.5\%$ of the model, but because a dense $1{\times}1$ convolution runs at full input resolution it costs $\sim\!0.57$\,GFLOP at $256^2$, about $1.0\%$ of the model's total forward FLOPs (Section~\ref{sec:limitations}).

\subsection{Laplacian-pyramid skips and high-frequency blocks}
\label{sec:laplacian}
Instead of max-pooling, each encoder junction performs a Laplacian-pyramid split. Given the level input $z$, a fixed Gaussian blur ($5{\times}5$) and stride-2 blur-pool produce the low band $\ell = \mathrm{blurpool}(\mathrm{gauss}(z))$, and the high band is the full-resolution residual $h = z - \mathrm{up}(\ell)$. The split is lossless ($\mathrm{up}(\ell)+h = z$) and channel-preserving. The low band $\ell$ descends to the next level. The high band $h$ becomes the skip connection, but only after passing through three additional bias-free \convnext{}-V1 blocks (\texttt{skip\_highfreq}), applied at both encoder levels. This is a wavelet-shrinkage-style inductive bias. Gaussian noise is spectrally flat while natural-image structure concentrates at low frequencies, so forcing the decoder to reconstruct from an explicit low-band descent plus a processed high-band residual denies it the trivial full-resolution identity-copy shortcut and matches the signal/noise spectral structure of the task.

\subsection{The \convnext{}-V1 block}
\label{sec:block}
Each block is a residual branch with call order
\[
x \;\mapsto\; x + \gamma \odot \Big[\, W_{\downarrow}\,\phi\big(\,W_{\uparrow}\,\mathrm{N}\big(\mathrm{DW}_{7\times7}\,x\big)\big)\Big],
\]
where $\mathrm{DW}_{7\times7}$ is a depthwise $7{\times}7$ convolution, $\mathrm{N}$ is the normalization (Section~\ref{sec:biasfree}), $W_{\uparrow}$ is a $1{\times}1$ expansion to $4\times$ width, $\phi$ is LeakyReLU with negative slope $0.1$ (followed by dropout $0.1$ at training time), $W_{\downarrow}$ is a $1{\times}1$ reduction, and $\gamma$ is a per-channel LayerScale multiplier (initialized $10^{-4}$, floored at $10^{-6}$). All convolutions are bias-free. We use the V1 block rather than V2, because the V2 variant inserts a Global Response Normalization whose learnable additive $\beta$ is bias-like and would break strict bias-freedom.

\subsection{Bias-free construction and degree-1 homogeneity}
\label{sec:biasfree}
The defining property of \model{} is that the entire map $D$ is positively homogeneous of degree one at inference:
\begin{equation}
D(\alpha y) = \alpha\, D(y) \qquad \text{for all } \alpha > 0.
\label{eq:homog}
\end{equation}
This holds because every constituent operation is linear-through-the-origin or degree-1 homogeneous, and no operation injects an additive constant:
\begin{itemize}
  \item every convolution (Gabor stem, $1{\times}1$ mix, all block convolutions, the head) has \texttt{use\_bias=False};
  \item the head activation is linear ($f(x)=x$);
  \item LeakyReLU is positively homogeneous, $\mathrm{LReLU}(\alpha x)=\alpha\,\mathrm{LReLU}(x)$ for $\alpha>0$ (this is why LeakyReLU is used rather than GELU, which is not homogeneous);
  \item normalization is a \emph{variance-only} batch norm, \bfbn{}: it has no running mean and no $\beta$ offset, only a fixed non-trainable running variance, so at inference $\mathrm{N}(x) = \gamma\, x / \sqrt{\mathrm{Var}+\epsilon}$ is exactly linear in $x$;
  \item the Laplacian split, bilinear upsampling, and additive skip merge are all linear.
\end{itemize}
The choice of \bfbn{} over the \convnext{} default LayerNorm is not cosmetic. LayerNorm divides by a per-sample statistic that itself scales with the input, making it scale-invariant (degree-0), which would destroy Eq.~\eqref{eq:homog} for the whole network. Homogeneity is an inference-time property: during training \bfbn{} uses the per-batch variance and is therefore degree-0, so Eq.~\eqref{eq:homog} holds only once the running statistics are frozen.

% ============================================================================
\section{Why bias-free: Miyasawa/Tweedie, and its limits}
\label{sec:theory}

Bias-freedom is more than an aesthetic choice: it connects the trained denoiser to a score, and it lets one blind model span a wide noise range. We give the motivation first, then its boundaries.

\paragraph{Miyasawa/Tweedie.}
For $y = x + \varepsilon$ with $\varepsilon \sim \mathcal{N}(0,\sigma^2 I)$ and marginal density $p_\sigma(y) = \int \mathcal{N}(y; x, \sigma^2 I)\, p(x)\, dx$, the minimum-MSE estimator obeys the Miyasawa/Tweedie identity
\begin{equation}
\hat{x}(y) \;=\; \E[x \mid y] \;=\; y + \sigma^2\, \nabla_y \log p_\sigma(y).
\label{eq:tweedie}
\end{equation}
Hence for an MMSE-optimal $D$ the denoiser residual $f(y) = D(y) - y$ is exactly $\sigma^2$ times the score $\nabla_y \log p_\sigma(y)$ of the noise-blurred density. A plain mean-squared-error objective is therefore well motivated: it drives $D$ toward the MMSE estimator whose residual is a scaled score. This is why \model{} is trained with MSE and read as an implicit-prior score model, with the caveat (made precise below) that the \emph{learned} residual realizes this only as a \emph{local} score estimate, not the gradient of a global density.

\paragraph{Homogeneity buys blind generalization.}
\citet{mohan2020biasfree} observe that a bias-free denoiser is degree-1 homogeneous (Eq.~\eqref{eq:homog}) and therefore behaves as a locally linear, input-adaptive filter $D(y) \approx J(y)\,y$ whose Jacobian $J(y)=\partial D/\partial y$ is degree-0 (scale-invariant). Because $J$ depends only on the direction of $y$, not its magnitude, a network trained over one range of noise levels extrapolates gracefully to unseen levels. A biased network, whose additive terms fix an absolute scale, instead collapses off its training range. This is the property that makes a single blind model viable, and it is the empirical target of our noise curriculum (Section~\ref{sec:training}).

\paragraph{The limits of the homogeneity argument.}
\label{sec:theory-limits}
Eq.~\eqref{eq:tweedie} is easy to over-read, in two ways.
\begin{enumerate}
  \item \textbf{Homogeneity is checkpoint-specific, not an architectural guarantee.} A network is exactly homogeneous only if every operation is genuinely bias-free. For this checkpoint the relative homogeneity error $\lVert D(\alpha y)-\alpha D(y)\rVert / \lVert \alpha D(y)\rVert$ is exactly $0$ across $\alpha \in \{\tfrac14,\tfrac12,2,4,8\}$ (these are powers of two, for which the scaling is exact in binary floating point), and it stays at the level of floating-point round-off, below $1\times10^{-4}$, under a non-dyadic stress test at $\alpha \in \{0.3,1.3,3.7,6.1\}$. The property is fragile: a single non-bias-free operation, such as a LayerNorm, drives the homogeneity error up by four orders of magnitude, to order $\sim\!0.9$. Eq.~\eqref{eq:homog} must therefore be verified per checkpoint, not assumed.
  \item \textbf{The learned residual is not a conservative field.} Eq.~\eqref{eq:tweedie} would make $f(y)$ the gradient of the scalar $\sigma^2 \log p_\sigma(y)$ only if the Jacobian $J(y)$ were symmetric, and it is not. For this checkpoint the measured Jacobian asymmetry $\lVert J - J^\top\rVert / \lVert J\rVert$ is $0.93$, roughly three orders of magnitude ($\sim\!2400\times$) above a symmetric-blur control, so no global energy or log-density function exists whose gradient is the residual. Plug-and-play and Regularization-by-Denoising convergence guarantees assume a conservative, symmetric-Jacobian denoiser, so they do not transfer here, and the residual should be read as a useful local score estimate rather than a calibrated global prior.
\end{enumerate}
We return to the practical consequences in Section~\ref{sec:limitations}.

% ============================================================================
\clearpage
\section{Training}
\label{sec:training}

\paragraph{Data and noise curriculum.}
We train on $256{\times}256$ patches drawn from COCO (\texttt{train2017}) and DIV2K (\texttt{train}), sourced with a per-directory weighting so the $\sim$118K COCO images do not drown DIV2K's $\sim$800. Patches are normalized to $[0,1]$ (peak-to-peak range $1.0$) and corrupted with additive Gaussian noise whose per-patch standard deviation is drawn uniformly from $[0,\, \sigma_{\max}]$, where the upper bound $\sigma_{\max}$ is widened linearly over training by a curriculum callback from $0.025$ to $0.25$ across $100$ epochs. Starting narrow and widening stabilizes early optimization, and bias-freedom then supplies the cross-level generalization. The noisy input is clipped to $[0,1]$ immediately after the noise is added, so no pixel is ever presented to the network above $1$ or below $0$; the identical clip is applied at inference, so the model operates on the same bounded domain in both phases. (At large $\sigma$ this makes the effective corruption a clipped Gaussian rather than a purely additive one.)

\paragraph{The $\sigma$ convention.}
The model operates entirely in $[0,1]$, so its noise level $\sigma_{\mathrm{norm}} \in [0, 0.25]$. The literature reports noise on the $[0,255]$ pixel scale, and the two are related by $\sigpx = \sigma_{\mathrm{norm}} \cdot 255$. The curriculum's $\sigma_{\max}: 0.025 \to 0.25$ therefore spans $\sigpx \approx 6.4 \to 64$, which covers the classic $15/25/50$ benchmark regimes as a single blind model. All PSNR values are computed with $\mathrm{max\_val}=1.0$ (matched to the peak-to-peak range). Because PSNR is scale-invariant under a matched range, these dB numbers are directly comparable to published $\mathrm{max\_val}=255$ results, with no further conversion.

\paragraph{Optimization.}
Loss is plain MSE (the Miyasawa-optimal objective, Section~\ref{sec:theory}). We use AdamW (learning rate $10^{-3}$, cosine decay, $10$ warmup epochs, decoupled weight decay $4\times10^{-3}$), gradient-norm clipping at $1.0$, batch size $4$, $100$ epochs.

% ============================================================================
\clearpage
\section{Results}
\label{sec:results}

We report two evaluations of the single trained checkpoint. (i) A blind PSNR-vs-noise sweep on DIV2K-validation ($100$ patches at $256$px). (ii) The standard leaderboard comparison on all four color sets (CBSD68, Kodak24, McMaster, Urban100) at $\sigma \in \{15,25,50\}$ using the full-image protocol (whole images, reflect-padded to a multiple of $16$), the same protocol the published numbers use. One caveat applies: our inputs are clipped to $[0,1]$ (Section~\ref{sec:training}), whereas the collated baselines follow the more common unclipped-AWGN convention, and we quantify the effect of this asymmetry in Section~\ref{sec:limitations}.

\subsection{Blind denoising across noise levels}
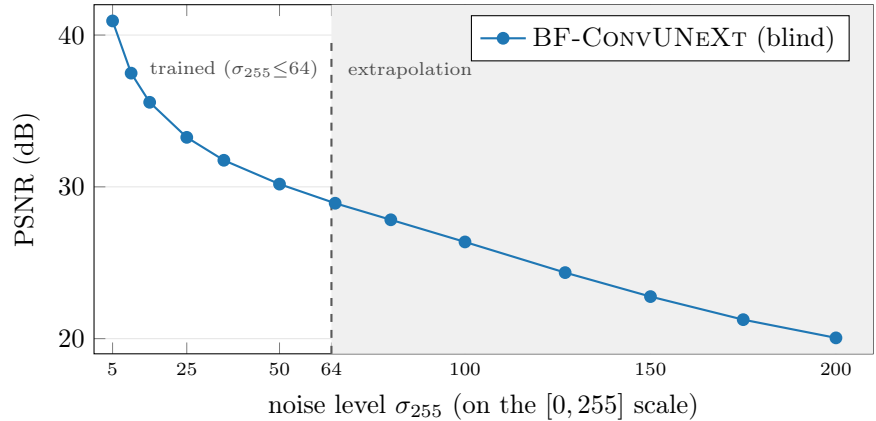
\begin{figure}[H]
\centering
\begin{tikzpicture}
\begin{axis}[
  width=0.72\textwidth, height=6.2cm,
  xlabel={noise level $\sigpx$ (on the $[0,255]$ scale)},
  ylabel={PSNR (dB)},
  xmin=0, xmax=210, ymin=19, ymax=42,
  xtick={5,25,50,64,100,150,200},
  xticklabel style={font=\scriptsize},
  ymajorgrids=true, grid style={gray!20},
  legend pos=north east, legend cell align={left},
  mark size=2pt,
]
% shade the extrapolation region (beyond the training max sigma_norm=0.25, i.e. sigma_255~64)
\addplot[draw=none, fill=gray!12, forget plot] coordinates {(64,19) (210,19) (210,42) (64,42)} \closedcycle;
\addplot[oursblue, thick, mark=*] coordinates {
  (5,40.93) (10,37.49) (15,35.57) (25,33.26) (35,31.75) (50,30.18) (65,28.92) (80,27.83) (100,26.37) (127,24.35) (150,22.77) (175,21.25) (200,20.05)
};
\addlegendentry{\model{} (blind)}
\draw[dashed, gray!70!black, thick] (axis cs:64,19) -- (axis cs:64,40);
\node[font=\scriptsize, gray!60!black, anchor=south east] at (axis cs:63,36.5) {trained ($\sigpx{\le}64$)};
\node[font=\scriptsize, gray!60!black, anchor=south west] at (axis cs:66,36.5) {extrapolation};
\end{axis}
\end{tikzpicture}
\caption{Blind PSNR vs.\ noise on DIV2K-validation ($100$ patches, $256$px). One model, no noise-level input, across the full range and \textbf{beyond it}: the curriculum trained only up to $\sigpx{\approx}64$ (dashed line), yet the model keeps denoising in the shaded extrapolation region: $22.8$\,dB at $\sigpx{=}150$ and $20.0$\,dB at $\sigpx{=}200$ ($2.3\times$ and $3.1\times$ the training maximum), degrading smoothly with no cliff. This graceful off-range behaviour is the signature of the bias-free/degree-1-homogeneous construction (Section~\ref{sec:theory}); a biased network collapses once past its training noise levels. The $95\%$ bootstrap CIs (not shown) are $\pm 0.4$--$0.8$\,dB.}
\label{fig:psnrsweep}
\end{figure}

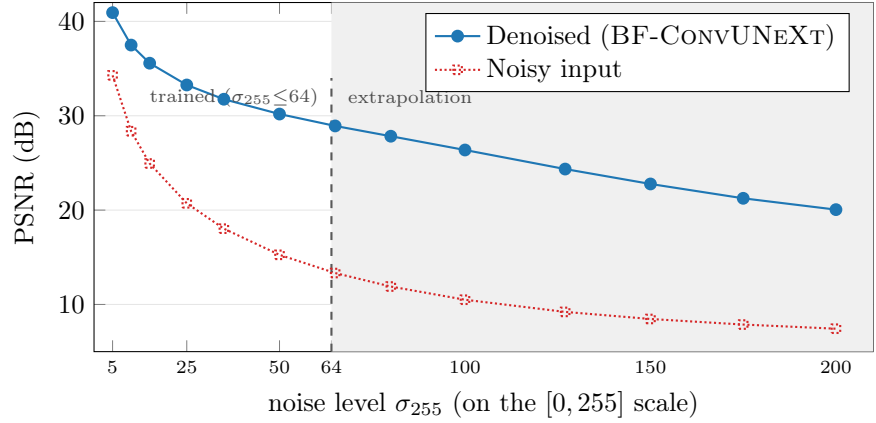
\begin{figure}[H]
\centering
\begin{tikzpicture}
\begin{axis}[
  width=0.72\textwidth, height=6.2cm,
  xlabel={noise level $\sigpx$ (on the $[0,255]$ scale)},
  ylabel={PSNR (dB)},
  xmin=0, xmax=210, ymin=5, ymax=42,
  xtick={5,25,50,64,100,150,200},
  xticklabel style={font=\scriptsize},
  ymajorgrids=true, grid style={gray!20},
  legend pos=north east, legend cell align={left},
  mark size=2pt,
]
% shade the extrapolation region beyond the training max (sigma_norm=0.25, i.e. sigma_255~64)
\addplot[draw=none, fill=gray!12, forget plot] coordinates {(64,5) (210,5) (210,42) (64,42)} \closedcycle;
\addplot[oursblue, thick, mark=*] coordinates {
  (5,40.93) (10,37.49) (15,35.57) (25,33.26) (35,31.75) (50,30.18) (65,28.92) (80,27.83) (100,26.37) (127,24.35) (150,22.77) (175,21.25) (200,20.05)
};
\addlegendentry{Denoised (\model{})}
\addplot[skipred, thick, densely dotted, mark=square, mark size=1.6pt] coordinates {
  (5,34.28) (10,28.36) (15,24.94) (25,20.72) (35,18.03) (50,15.26) (65,13.33) (80,11.90) (100,10.50) (127,9.21) (150,8.46) (175,7.87) (200,7.43)
};
\addlegendentry{Noisy input}
\draw[dashed, gray!70!black, thick] (axis cs:64,5) -- (axis cs:64,34);
\node[font=\scriptsize, gray!60!black, anchor=south east] at (axis cs:63,30) {trained ($\sigpx{\le}64$)};
\node[font=\scriptsize, gray!60!black, anchor=south west] at (axis cs:66,30) {extrapolation};
\end{axis}
\end{tikzpicture}
\caption{Denoising gain: denoised PSNR (solid blue) vs.\ the corrupted noisy-input PSNR (dotted red) across the noise sweep on DIV2K-validation ($100$ patches, $256$px). The vertical gap between the two curves is the per-level PSNR gain from a single blind model: $\sim\!6.7$\,dB at $\sigpx{=}5$, widening to $\sim\!15.6$--$15.9$\,dB around $\sigpx{=}65$--$80$. The gain persists far past the $\sigpx{\approx}64$ training ceiling (dashed line; shaded extrapolation region), still $+12.6$\,dB at $\sigpx{=}200$ ($20.0$ vs.\ $7.4$\,dB), consistent with the graceful off-range behaviour in Figure~\ref{fig:psnrsweep}.}
\label{fig:gain}
\end{figure}

Figure~\ref{fig:psnrsweep} shows the blind PSNR-vs-noise sweep on DIV2K-validation, where bias-freedom matters most. A single model, given no noise-level input, denoises across the whole curriculum range ($40.9$\,dB at $\sigpx{=}5$ to $28.9$\,dB at $\sigpx{=}65$, just past the $\sigpx{\approx}64$ training maximum), and it keeps working \emph{far beyond} that maximum. The curriculum only ever showed the network $\sigma_{\mathrm{norm}}\!\le\!0.25$; evaluated at $\sigma_{\mathrm{norm}}{=}0.59$ and $0.78$ ($\sigpx{=}150$ and $200$, $2.3\times$ and $3.1\times$ the largest training noise) it still returns $22.8$ and $20.0$\,dB, degrading smoothly and monotonically with no cliff even at more than three times the training ceiling. This is the extrapolation predicted by the degree-1 homogeneity of Section~\ref{sec:theory}, which \citet{mohan2020biasfree} argue a \emph{biased} network cannot manage. (Gains are quoted against the noisy input \emph{as fed to the model}, clipped to the $[0,1]$ domain, which at high $\sigma$ has a higher PSNR than the unclipped Gaussian.) Figure~\ref{fig:gain} plots the same sweep against the noisy-input baseline, making the per-level denoising gain explicit.

\subsection{Comparison against the color-denoising leaderboard}
Table~\ref{tab:sota} places \model{} against six published baselines on all four standard color sets at $\sigma\in\{15,25,50\}$, under the identical full-image protocol and PSNR convention. The single $0.82$M blind model matches or beats DnCNN and FFDNet on every set at every noise level, and trails the heavyweight state of the art by a small, consistent margin that widens on Urban100.

\begin{table}[H]
\centering
\caption{Color AWGN denoising, PSNR (dB), full-image protocol, at $\sigma\in\{15,25,50\}$. Baseline numbers are the published values collated by \citet{zhang2022scunet}. \textbf{Bold} marks where \model{} beats a given baseline. Approximate parameter counts are listed in the header; DnCNN/FFDNet are $<1$M, the four rightmost are $12$--$32$M.}
\label{tab:sota}
\renewcommand{\arraystretch}{1.12}
\resizebox{\textwidth}{!}{%
\begin{tabular}{llccccccc}
\toprule
\textbf{Set} & \textbf{$\sigpx$} & \textbf{\model{}} & \textbf{DnCNN} & \textbf{FFDNet} & \textbf{DRUNet} & \textbf{SwinIR} & \textbf{Restormer} & \textbf{SCUNet} \\
 & & \textbf{(0.82M)} & ($0.6$M) & ($0.5$M) & ($32$M) & ($12$M) & ($26$M) & ($18$M) \\
\midrule
CBSD68   & 15 & 34.00 & \textbf{33.90} & \textbf{33.87} & 34.30 & 34.42 & 34.40 & 34.40 \\
CBSD68   & 25 & 31.35 & \textbf{31.24} & \textbf{31.21} & 31.69 & 31.78 & 31.79 & 31.79 \\
CBSD68   & 50 & 28.10 & \textbf{27.95} & \textbf{27.96} & 28.51 & 28.56 & 28.60 & 28.61 \\
\midrule
Kodak24  & 15 & 34.99 & \textbf{34.60} & \textbf{34.63} & 35.31 & 35.34 & 35.47 & 35.34 \\
Kodak24  & 25 & 32.51 & \textbf{32.14} & \textbf{32.13} & 32.89 & 32.89 & 33.04 & 32.92 \\
Kodak24  & 50 & 29.35 & \textbf{28.95} & \textbf{28.98} & 29.86 & 29.79 & 30.01 & 29.87 \\
\midrule
McMaster & 15 & 34.86 & \textbf{33.45} & \textbf{34.66} & 35.40 & 35.61 & 35.61 & 35.60 \\
McMaster & 25 & 32.59 & \textbf{31.52} & \textbf{32.35} & 33.14 & 33.20 & 33.34 & 33.34 \\
McMaster & 50 & 29.44 & \textbf{28.62} & \textbf{29.18} & 30.08 & 30.22 & 30.30 & 30.29 \\
\midrule
Urban100 & 15 & 34.24 & \textbf{32.98} & \textbf{33.83} & 34.81 & 35.13 & 35.13 & 35.18 \\
Urban100 & 25 & 31.85 & \textbf{30.81} & \textbf{31.40} & 32.60 & 32.90 & 32.96 & 33.03 \\
Urban100 & 50 & 28.50 & \textbf{27.59} & \textbf{28.05} & 29.61 & 29.82 & 30.02 & 30.14 \\
\midrule
\multicolumn{2}{l}{\emph{avg} $\sigma{=}15$} & 34.52 & \textbf{33.73} & \textbf{34.25} & 34.96 & 35.13 & 35.15 & 35.13 \\
\multicolumn{2}{l}{\emph{avg} $\sigma{=}25$} & 32.07 & \textbf{31.43} & \textbf{31.77} & 32.58 & 32.69 & 32.78 & 32.77 \\
\multicolumn{2}{l}{\emph{avg} $\sigma{=}50$} & 28.85 & \textbf{28.28} & \textbf{28.54} & 29.52 & 29.60 & 29.73 & 29.73 \\
\bottomrule
\end{tabular}%
}
\end{table}

Averaged over the four sets, \model{} is $+0.79/+0.64/+0.57$\,dB above DnCNN at $\sigma=15/25/50$ and $+0.27/+0.30/+0.31$\,dB above FFDNet (which, unlike our model, is \emph{non-blind} and receives a noise-level map). It is $-0.44/-0.51/-0.67$\,dB relative to DRUNet and $-0.63/-0.71/-0.88$\,dB relative to Restormer, the leaderboard top. \model{} is ahead of DnCNN in all twelve cells, but the per-set margin is uneven: it is narrow on CBSD68 ($+0.10$ to $+0.15$\,dB) and much larger on McMaster and Urban100 (up to $+1.41$ and $+1.26$\,dB at $\sigma{=}15$). On CBSD68 that margin is within the sensitivity of our $[0,1]$ input clip, so the result there is best read as a match rather than a clear win (Section~\ref{sec:limitations}); the McMaster, Urban100, and Kodak24 margins are far larger. Against the heavyweight state of the art the gap is smallest on CBSD68 ($\sim0.3$--$0.5$\,dB), then Kodak24 ($\sim0.3$--$0.7$\,dB) and McMaster ($\sim0.5$--$0.9$\,dB), and largest on Urban100 ($\sim0.6$--$1.7$\,dB). Urban100's repetitive, self-similar structures are exactly where the transformers' non-local attention (SwinIR, Restormer) and DRUNet's large receptive field have the most to exploit and a compact local conv U-Net the least.

\subsection{Structural similarity (SSIM)}
\label{sec:ssim}
Table~\ref{tab:ssim} reports SSIM alongside PSNR for \model{} on the same four sets and full-image protocol (same checkpoint as Table~\ref{tab:sota}), computed on the $[0,1]$ representation with a data range of $1.0$. SSIM tracks PSNR's ordering across noise, highest at $\sigpx{=}15$ (average $0.9318$) and degrading gracefully to average $0.8221$ at $\sigpx{=}50$. Urban100 is a cross-metric exception: it has the highest SSIM of the four sets at every noise level ($0.945/0.918/0.857$) despite its middling PSNR. Its regular, high-contrast man-made structures, the same self-similarity that costs PSNR against non-local models, yield strong local structural agreement once denoised. We report SSIM for \model{} only, because our comparison source \citep{zhang2022scunet} tabulates PSNR alone and SSIM values assembled from mixed sources use inconsistent windowing, border, and color conventions, so a baseline SSIM column would be an unreliable comparison.

\begin{table}[H]
\centering
\caption{\model{} PSNR (dB) and SSIM on the four color benchmark sets, full-image protocol, single blind checkpoint (identical protocol and model as Table~\ref{tab:sota}; the PSNR columns reproduce it exactly). SSIM on $[0,1]$, data range $1.0$. Baseline SSIM is not tabulated (see Section~\ref{sec:ssim}).}
\label{tab:ssim}
\renewcommand{\arraystretch}{1.15}
\begin{tabular}{lcccccc}
\toprule
 & \multicolumn{2}{c}{$\sigpx=15$} & \multicolumn{2}{c}{$\sigpx=25$} & \multicolumn{2}{c}{$\sigpx=50$} \\
\cmidrule(lr){2-3}\cmidrule(lr){4-5}\cmidrule(lr){6-7}
\textbf{Set} & PSNR & SSIM & PSNR & SSIM & PSNR & SSIM \\
\midrule
CBSD68   & 34.00 & 0.9316 & 31.35 & 0.8870 & 28.10 & 0.7981 \\
Kodak24  & 34.99 & 0.9259 & 32.51 & 0.8855 & 29.35 & 0.8061 \\
McMaster & 34.86 & 0.9241 & 32.59 & 0.8918 & 29.44 & 0.8269 \\
Urban100 & 34.24 & 0.9454 & 31.85 & 0.9178 & 28.50 & 0.8572 \\
\midrule
\emph{average} & 34.52 & 0.9318 & 32.07 & 0.8955 & 28.85 & 0.8221 \\
\bottomrule
\end{tabular}
\end{table}

\subsection{Qualitative examples}
\label{sec:qualitative}
\begin{figure}[H]
\centering
\includegraphics[width=\linewidth]{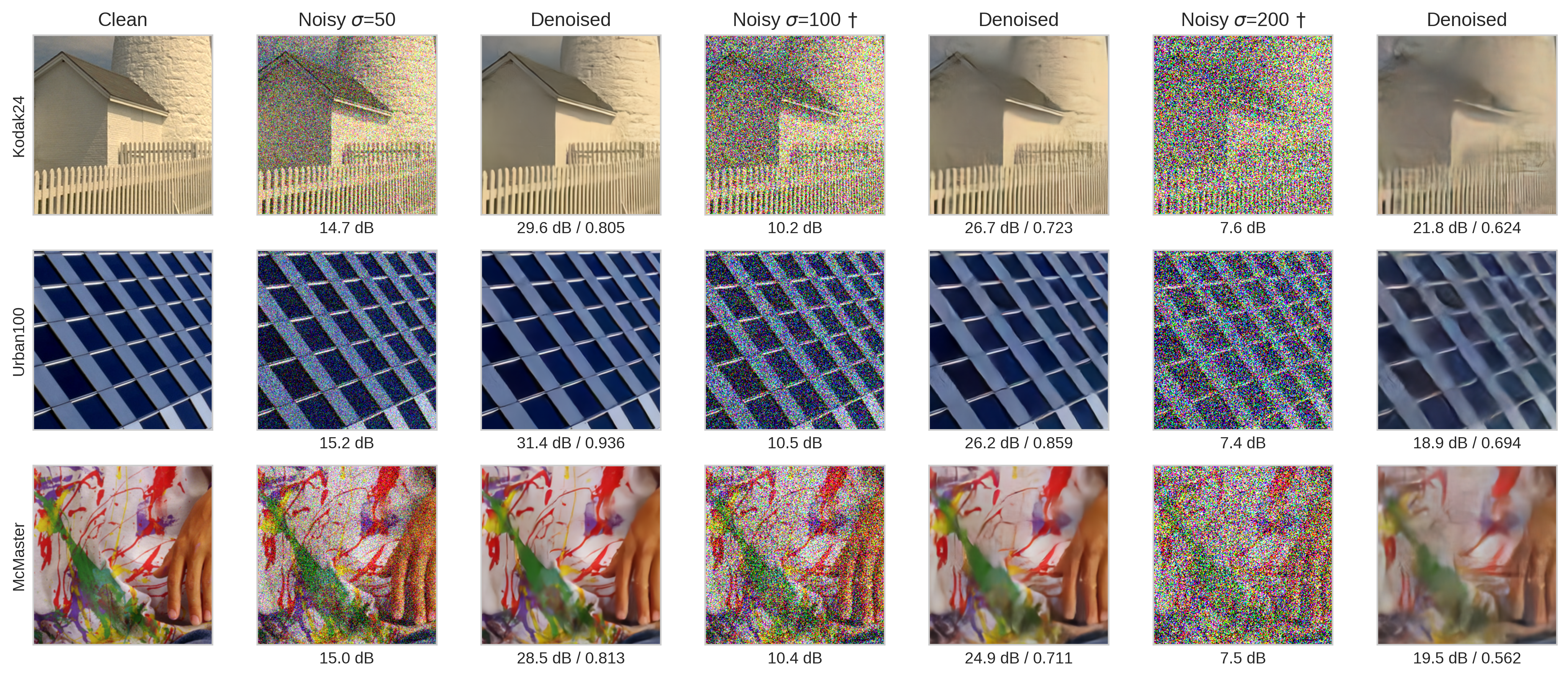}
\caption{A clean reference followed by (noisy, denoised) pairs at $\sigpx\in\{50,100,200\}$ on three held-out benchmark images (rows: Kodak24, Urban100, McMaster). One frozen blind model, no noise-level input, at every level. $\dagger$ marks $\sigpx\in\{100,200\}$, beyond the $\sigpx{\le}64$ training range: from about $7.5$\,dB of static at $\sigpx{=}200$ the model still returns a $19$--$22$\,dB image. Per-image PSNR and SSIM (full image, $\mathrm{max\_val}{=}1$) are annotated under each panel.}
\label{fig:qualitative}
\end{figure}

Figure~\ref{fig:qualitative} shows the model on three held-out benchmark images (a Kodak24 scene, an Urban100 facade with fine repetitive structure that is a known denoiser stress case, and a McMaster crop with saturated color) across a wide sweep of noise, $\sigpx\in\{50,100,200\}$, from moderate corruption to near-total static. The highest level, $\sigpx{=}200$, is $3.1\times$ the training ceiling ($\sigpx{\le}64$); the same frozen blind model is applied at every level with no noise-level input. Full-image inference is run under the paper's protocol; the panels are a detail crop of the result. Even where the input is almost pure noise, and outside the trained range, edges and repeated structure come back cleanly, matching the out-of-range curve in Figure~\ref{fig:psnrsweep}. Per-image PSNR/SSIM (computed on the full image) are annotated.

\subsection{Accuracy for the parameter budget}
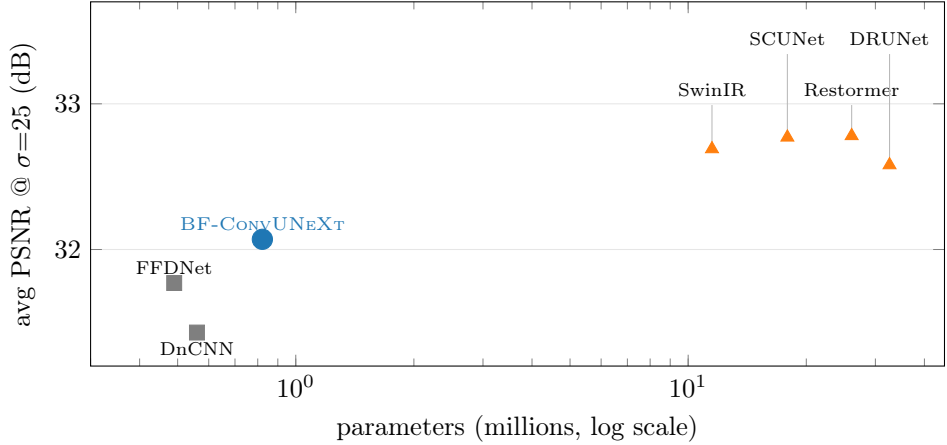
\begin{figure}[H]
\centering
\begin{tikzpicture}
\begin{axis}[
  width=0.78\textwidth, height=6.4cm,
  xlabel={parameters (millions, log scale)},
  ylabel={avg PSNR @ $\sigma{=}25$ (dB)},
  xmode=log, log basis x=10,
  xmin=0.3, xmax=45, ymin=31.2, ymax=33.7,
  ymajorgrids=true, grid style={gray!20},
  only marks, mark size=2.8pt,
]
\addplot[classicgray, mark=square*] coordinates {(0.49,31.77) (0.56,31.43)};
\addplot[oursblue, mark=*, mark size=3.8pt] coordinates {(0.822,32.07)};
\addplot[sotaorange, mark=triangle*] coordinates {(11.5,32.69) (17.9,32.77) (26.1,32.78) (32.6,32.58)};
% manual, collision-free point labels
\node[font=\scriptsize, anchor=south]                 at (axis cs:0.49,31.77) {FFDNet};
\node[font=\scriptsize, anchor=north]                 at (axis cs:0.56,31.43) {DnCNN};
\node[font=\scriptsize, anchor=south, text=oursblue]  at (axis cs:0.822,32.07)  {\model{}};
% SOTA points cluster tightly on the log axis; label them above with thin leaders
\node[font=\scriptsize] (lswin) at (axis cs:11.5,33.10) {SwinIR};
\node[font=\scriptsize] (lscu)  at (axis cs:17.9,33.45) {SCUNet};
\node[font=\scriptsize] (lrest) at (axis cs:26.1,33.10) {Restormer};
\node[font=\scriptsize] (ldru)  at (axis cs:32.6,33.45) {DRUNet};
\draw[gray!55, thin] (lswin) -- (axis cs:11.5,32.73);
\draw[gray!55, thin] (lscu)  -- (axis cs:17.9,32.81);
\draw[gray!55, thin] (lrest) -- (axis cs:26.1,32.82);
\draw[gray!55, thin] (ldru)  -- (axis cs:32.6,32.62);
\end{axis}
\end{tikzpicture}
\caption{Four-set average PSNR at $\sigma{=}25$ vs.\ (approximate) parameter count. \model{} (blue) sits on the efficient side of the trade-off: above the sub-1M CNNs and within $\sim0.71$\,dB of the $12$--$32$M state of the art at $\sim1/15$--$1/39$ of their size. Parameter counts are approximate and collated from the respective papers.}
\label{fig:params}
\end{figure}

The point of Table~\ref{tab:sota} is not that \model{} wins outright, because it does not, but where it sits on the accuracy/size trade-off. Figure~\ref{fig:params} plots the four-set average PSNR at $\sigma=25$ against approximate parameter count. \model{} lands between the two size classes: comfortably above the sub-1M classic CNNs it beats, and within about a dB of models $15$--$39\times$ larger.

The parameter count is only half the efficiency story; Table~\ref{tab:efficiency} adds compute and memory, measured on one RTX~4090. At $256^2$ the model is $\sim59$\,GFLOPs and peaks at $\sim582$\,MiB of GPU memory; FLOPs and memory both scale with pixel count (a fully-convolutional model), so a $512^2$ image is exactly $4\times$ the FLOPs and $\sim4.0\times$ the peak memory. Wall-clock latency, however, does not track pixel count at batch~$1$: a single-image forward pass takes $\sim0.12$\,s at both resolutions ($118$ and $118$\,ms median, indistinguishable despite $4\times$ the compute). At batch~$1$ the forward pass is dominated by kernel-dispatch and launch overhead rather than by compute, so it sits on a latency floor, and batched throughput is substantially higher. The ``compact'' claim is about parameters, FLOPs, and memory, and does not lean on any low-precision trick.

\begin{table}[H]
\centering
\caption{Efficiency of \model{}, measured on one RTX~4090 (single-image forward pass). FLOPs are per forward pass; memory is the peak GPU allocation. Cost is linear in pixel count. Latency is not tabulated because at batch~$1$ it sits on a dispatch-overhead floor rather than tracking compute (see text).}
\label{tab:efficiency}
\renewcommand{\arraystretch}{1.15}
\begin{tabular}{lcccc}
\toprule
Resolution & Parameters & FLOPs & FLOPs/megapixel & Peak GPU mem \\
\midrule
$256\times256$ & $0.82$M & $59.26$\,GFLOP & $904.18$\,GFLOP & $581.7$\,MiB \\
$512\times512$ & $0.82$M & $237.02$\,GFLOP & $904.18$\,GFLOP & $2312.2$\,MiB \\
\bottomrule
\end{tabular}
\end{table}

% ============================================================================
\section{Discussion: limitations and what does not work}
\label{sec:limitations}

\paragraph{Where the model stands.}
As a compact, blind denoiser \model{} is strong: it matches or beats the classic sub-1M CNNs everywhere (a clear win on three of the four sets; a match on CBSD68 within the input-clip protocol margin) and closes most of the gap to models an order of magnitude larger, from a single checkpoint with no noise-level input. It does not beat the heavyweight state of the art, and the gap is structurally largest on Urban100, where non-local self-similarity rewards the attention and large receptive fields that a two-level local conv U-Net lacks.

\paragraph{Why parameter count understates cost.}
Two of \model{}'s layers illustrate why a ``compact'' claim should not rest on parameters alone. The trainable stem projection (Section~\ref{sec:arch}) and the dense $1{\times}1$ head are both $1{\times}1$ convolutions evaluated at \emph{full input resolution}, where a one-pixel kernel still costs one multiply--accumulate per output pixel per channel pair. Together they are $4{,}554$ parameters, $0.6\%$ of the model, but $\sim\!0.6$\,GFLOP at $256^2$, or $\sim\!1.0\%$ of the model's total forward FLOPs --- roughly a two-fold gap between the two framings of the same two layers, and the gap widens with resolution while the parameter share stays fixed. We therefore report FLOPs and peak memory alongside parameters in Table~\ref{tab:efficiency}, and read the parameter column as the least informative of the three. The latency column is omitted deliberately: at batch~$1$ the forward pass is dispatch-bound, so the $512^2$ median is not larger than the $256^2$ median despite $4\times$ the compute, and quoting it would misrepresent the cost model.

\paragraph{Limits of the theory.}
The one genuinely negative result is the theory limit of Section~\ref{sec:theory-limits}: the learned residual is a local score, not the gradient of a global density, so uses that require a conservative denoiser (provable plug-and-play/RED convergence, energy-based sampling toward a fixed target, a calibrated Bayesian posterior) are out of reach. The homogeneity that does hold is inference-only and must be re-verified per checkpoint.

\paragraph{What the local score still buys.}
That the residual is only a local score, not a global prior, is less limiting than it may appear. In the annealed coarse-to-fine stochastic ascent of \citet{kadkhodaie2021stochastic}, the denoiser residual is used directly as the score at each effective noise level ($y \leftarrow y + h_t\,d_t + \gamma_t z_t$ with $d_t = (I - MM^\top)\,f(y) + M(m - M^\top y)$ for a linear measurement operator $M$ with orthonormal columns), and this loop nowhere assumes a symmetric Jacobian or a global energy. The same frozen model, unchanged, therefore drives unconstrained sampling from the implicit prior (the $M{=}0$ case) and solves a family of linear inverse problems (inpainting, random-pixel recovery, super-resolution, spectral deblurring, compressive sensing) from one blind denoiser. Unlike plug-and-play/RED it offers no fixed-point convergence guarantee and needs no conservative operator; it is an empirical sampler. A local score, though weaker than a global prior, therefore still supports these downstream uses.

\paragraph{Measurement limits.}
We report PSNR and SSIM for \model{} (Tables~\ref{tab:sota} and \ref{tab:ssim}), but the baseline comparison is PSNR-only, because our collated source \citep{zhang2022scunet} reports PSNR and cross-source SSIM is protocol-inconsistent. Table~\ref{tab:sota} shows point PSNR without per-cell error bars: with a fixed noise seed each cell is a deterministic function of the frozen model, so the relevant uncertainty is the across-image spread, which the DIV2K sweep reports as $\pm0.4$--$0.8$\,dB $95\%$ bootstrap CIs (Figure~\ref{fig:psnrsweep}); these are indicative of, but not measured on, the leaderboard sets. All results are from a single checkpoint, and we present no ablations: the individual contributions of the frozen Gabor stem, the Laplacian high-frequency skips, the \texttt{high\_freq\_blocks}, the bias-free normalization, and the noise curriculum are not isolated. Our baselines are externally collated rather than run in-house; a bias-free flat CNN would supply the missing closest-competitor row and isolate the \emph{architecture}, though not bias-freedom itself, for which the clean control is this same network trained \emph{with} biases and a standard normalization. One protocol asymmetry deserves an explicit caveat: we clip the noisy input to $[0,1]$ at both training and inference (Section~\ref{sec:training}), whereas the standard AWGN benchmark convention is unclipped additive noise, and the collated baselines almost certainly follow it. Because clipping removes the out-of-range excursions that carry pure error, the clipped input is an easier starting point. We measured the size of this advantage directly, by re-running the CBSD68 evaluation of this same checkpoint under the identical full-image protocol with the clip disabled: it is worth $+0.28$, $+0.71$, and $+2.23$\,dB at $\sigma_{255}{=}15$, $25$, and $50$. This has two consequences. At every level the clip advantage exceeds our $+0.10$ to $+0.15$\,dB margin over DnCNN on CBSD68, so we read that comparison as a match rather than a win. It also grows steeply with $\sigma$, since clipping discards a larger share of the noise as the noise grows, and by $\sigma_{255}{=}50$ it is several times the $\sim0.3$--$0.5$\,dB gap we report to the heavyweight state of the art on that set. The effect inflates only our own numbers, so it never narrows that gap; but it does mean that at the highest level our CBSD68 figures are not directly comparable to unclipped-protocol numbers, and a like-for-like unclipped comparison there would place us further behind the leaderboard than Table~\ref{tab:sota} suggests, not closer. We measured this on CBSD68 only. The McMaster, Urban100, and Kodak24 margins over DnCNN/FFDNet are wide enough to absorb an effect of the $\sigma_{255}{=}15$--$25$ magnitude, but we have not measured the clip advantage on those sets and do not claim their $\sigma_{255}{=}50$ margins survive it. A perceptual metric such as LPIPS, and grayscale Set12/BSD68 (which needs a one-channel model), are likewise outside the present scope. These are deliberate boundaries of a single-checkpoint study.

% ============================================================================
\section{Related work}
\label{sec:related}

\paragraph{Bias-free denoising.}
\citet{mohan2020biasfree} introduced bias-free CNNs for blind denoising and established the degree-1-homogeneity argument we build on. The residual-as-score reading rests on the classical Miyasawa/Tweedie identity \citep{miyasawa1961,robbins1956}, which also underpins the implicit-prior sampling line of \citet{kadkhodaie2021stochastic}. \model{} applies the bias-free discipline end-to-end to a modern block and front end.

\paragraph{CNN and transformer denoisers.}
DnCNN \citep{zhang2017dncnn} established the residual-CNN denoiser; FFDNet \citep{zhang2018ffdnet} added a noise-level map for non-blind flexibility; DRUNet \citep{zhang2021plugandplay} coupled a U-Net denoiser with plug-and-play restoration; SwinIR \citep{liang2021swinir}, Restormer \citep{zamir2022restormer}, and SCUNet \citep{zhang2022scunet} brought (windowed/channel) attention and are the current leaderboard leaders. These are our baselines; all except our model and DnCNN's blind variant either use a noise map or are larger by $15$--$39\times$.

\paragraph{Building blocks.}
The block is a \convnext{} \citep{liu2022convnext} in a U-Net \citep{ronneberger2015unet}. The multi-scale skip decomposition is a Laplacian pyramid \citep{burt1983laplacian}, and the frozen front end is a Gabor filter bank \citep{gabor1946}. \model{} contributes their bias-preserving integration and its measurement.

% ============================================================================
\section{Reproducibility}
\label{sec:repro}

The entire stack is reproducible and open: the model, the training recipe, the noise curriculum, the full-image evaluation harness, and the implicit-prior inverse-problem solver of Section~\ref{sec:limitations} are all implemented in the \texttt{dl\_techniques} framework at \url{https://github.com/NikolasMarkou/dl_techniques}.

The evaluated checkpoint is \texttt{20260720\_convunext\_denoiser\_hfb3} ($821{,}832$ parameters total, $807{,}048$ trainable and $14{,}784$ frozen; depth $2$; flat width $66$; three \convnext{}-V1 blocks per scope plus three per high-frequency skip band; frozen $22$-filter, $11{\times}11$ Gabor stem followed by a trainable bias-free $1{\times}1$ stem projection; \bfbn{}; LeakyReLU($0.1$); linear dense head). It was trained with the recipe of Section~\ref{sec:training}. The full-image benchmark uses whole test images reflect-padded to a multiple of $16$, PSNR at $\mathrm{max\_val}=1.0$ on $[0,1]$, on the standard CBSD68 / Kodak24 / McMaster / Urban100 sets. All numbers in Table~\ref{tab:sota}, Table~\ref{tab:ssim} and Figure~\ref{fig:psnrsweep} are measured from that single checkpoint; baseline numbers are the published values as collated by \citet{zhang2022scunet}. The efficiency numbers (Table~\ref{tab:efficiency}) are single-image forward passes on one RTX~4090.

\paragraph{Gabor stem, in full.} The frozen stem is fully specified so it can be regenerated deterministically (given the grid and seed below; the exact per-axis \texttt{linspace} resolution and channel-ordering are fixed in the released code). It is a depthwise $11{\times}11$ convolution with depth multiplier $22$ (so $3{\times}22{=}66$ output channels), \texttt{use\_bias}=False, \texttt{trainable}=False. Each output channel is one real 2D Gabor wavelet $g(u,v)=\exp\!\big(-\tfrac{u_\theta^2+\gamma^2 v_\theta^2}{2s^2}\big)\cos\!\big(\tfrac{2\pi u_\theta}{\lambda}+\psi\big)$, with $(u_\theta,v_\theta)$ the coordinates rotated by $\theta$. The five parameters are swept on a regular \texttt{linspace} grid over the Özbulak--Ekenel Table-I ranges ($s\in[2,21]$, $\theta\in[0^\circ,360^\circ]$, $\lambda\in[8,100]$, $\gamma\in[0,300\%]$, $\psi\in[0^\circ,360^\circ]$), one distinct $(s,\theta,\lambda,\gamma,\psi)$ tuple per output channel, replicated identically across the three input channels, with global seed $42$. Because the bank is fixed, it contributes zero trainable parameters and requires no checkpointing beyond these constants.

\paragraph{Stem projection, in full.} The one trainable layer between the frozen bank and encoder L0 is \texttt{gabor\_\allowbreak stem\_\allowbreak projection}: a \texttt{Conv2D} with $66$ filters, kernel $1{\times}1$, stride $1$, no padding needed, \texttt{use\_bias}=False, \texttt{trainable}=True, linear activation, and no normalization. Its kernel is therefore a single $66\times66$ matrix ($4{,}356$ parameters), initialized with the framework default (Glorot uniform) under the same global seed $42$, and it is trained jointly with the rest of the network under the recipe of Section~\ref{sec:training}. It applies the same $66\times66$ mix at every spatial position, so it is exactly a learned linear recombination of the frozen Gabor responses, and, being bias-free and linear, it preserves degree-1 homogeneity. The final head is a bias-free dense $1{\times}1$ \texttt{Conv2D} back to $3$ channels ($198$ parameters, $\texttt{groups}=1$, linear activation).

% ============================================================================
\section{Conclusion}
\label{sec:conclusion}

We described \model{}, a compact bias-free \convnext{} U-Net that integrates a frozen Gabor stem, Laplacian-pyramid high-frequency skips, and end-to-end degree-1 homogeneity, and we measured it against the color-denoising leaderboard. As a single $0.82$M-parameter blind model it matches or beats the classic sub-1M CNNs on all four standard sets at all three noise levels (a match on CBSD68 once the input clip is accounted for, Section~\ref{sec:limitations}) and comes within about a dB of state-of-the-art models an order of magnitude larger, with the residual gap concentrated on self-similar imagery. Within its stated limits (a single, non-conservative checkpoint) it is a useful data point on how much denoising accuracy, and, through the implicit-prior score, how much downstream inverse-problem capability, is reachable from a small, strictly bias-free model.

% ============================================================================
\bibliographystyle{plainnat}

\end{document}